\documentclass[aip,reprint,8pt]{revtex4-1}

\usepackage{lipsum}
\usepackage{amsmath}
\usepackage{graphicx}
\usepackage{dcolumn}
\usepackage{bm}
\usepackage[utf8]{inputenc}
\usepackage[T1]{fontenc}
\usepackage{mathptmx}
\usepackage[dvipsnames]{xcolor}
\usepackage{verbatim}
\usepackage{wrapfig}
\usepackage{gensymb}
\usepackage{footmisc}

\draft 

\begin{document}

\title{Dynamic polarizability of macromolecules for single-molecule optical biosensing} 

{
\makeatletter
\def\frontmatter@thefootnote{%
 \altaffilletter@sw{\@fnsymbol}{\@fnsymbol}{\csname c@\@mpfn\endcsname}%
}%
\makeatother

\author{Larnii S. Booth}
\affiliation{ARC Centre for Engineered Quantum Systems (EQUS), School of Mathematics and Physics, The University of Queensland, Australia}

\author{Eloise V. Browne}
\affiliation{ARC Centre for Engineered Quantum Systems (EQUS), School of Mathematics and Physics, The University of Queensland, Australia}

\author{Nicolas P. Mauranyapin}
\affiliation{ARC Centre for Engineered Quantum Systems (EQUS), School of Mathematics and Physics, The University of Queensland, Australia}

\author{Lars S. Madsen}
\affiliation{ARC Centre for Engineered Quantum Systems (EQUS), School of Mathematics and Physics, The University of Queensland, Australia}

\author{Shelley Barfoot}
\affiliation{School of Chemistry and Molecular Biosciences, The University of Queensland, Australia}

\author{Alan Mark}
\affiliation{School of Chemistry and Molecular Biosciences, The University of Queensland, Australia}

\author{Warwick P. Bowen}
\email[email: ]{w.bowen@uq.edu.au}
\affiliation{ARC Centre for Engineered Quantum Systems (EQUS), School of Mathematics and Physics, The University of Queensland, Australia}

\date{\today}

\begin{abstract}
The structural dynamics of macromolecules is important for most microbiological processes, from protein folding to the origins of neurodegenerative disorders. Noninvasive measurements of these dynamics are highly challenging. Recently, optical sensors have been shown to allow noninvasive time-resolved measurements of the dynamic polarizability of single-molecules. Here we introduce a method to efficiently predict the dynamic polarizability from the atomic configuration of a given macromolecule. This provides a means to connect the measured dynamic polarizability to the  underlying structure of the molecule, and therefore to connect temporal measurements to structural dynamics. To illustrate the methodology we calculate the change in polarizability as a function of time based on conformations extracted from molecular dynamics simulations and using different conformations of motor proteins solved crystalographically. This allows us to quantify the magnitude of the changes in polarizablity due to thermal and functional motions.

\end{abstract}


\maketitle 

\section{Introduction}

Dielectric interactions between light and matter are widely used in biological sensing, with  applications ranging from phase contrast microscopy~\cite{sanderson2001phase} to dynamic light scattering~\cite{postnov2020dynamic,junger2018strong}. In recent years a significant focus has been to extend the applications of these interactions into the regime of single macromolecules, with particular interest in label-free identification and dynamical tracking~\cite{mauranyapin2017evanescent,baaske2014single,young2018quantitative}. This is challenging because dielectric interactions are exceedingly weak for particles that are smaller than the optical wavelength, with the cross-section of dipole scattering scaling as the particle radius to the power of six~\cite{jackson_classical_1999}. Nevertheless, several techniques have been developed that are able to resolve macromolecules with sizes down to a few nanometers, far below the scale of the optical wavelength. Figure~\ref{fig1} illustrates a number of these techniques, including optical cavity or plasmonic resonance enhanced sensors~\cite{swaim2013detection,Swaim2011Detection, Shopova2011Plasmonic,  Pang2012Optical,Dantham2013Label-free,zijlstra2012optical}, dark-field heterodyne microscopy~\cite{mauranyapin2017evanescent, mauranyapin2019quantum, jin20211,mitra2012real}, interferometric scattering microscopy~\cite{young2018quantitative,Ortega2014Label-Free,mosby2020myosin,piliarik2014direct}, and plasmonic optical traps\cite{Kotnala2014Double, AlBalushi2014Label-free, Lin2014Trapping}.

It has been shown that single-molecule sensors which use dielectric interactions provide a signal that is highly correlated to the mass of the macromolecule being measured, independent of the molecular structure~\cite{vollmer2002protein,young2018quantitative}. This enables a powerful approach for analysing molecules, molecular assembly, and chemical reactions, termed {\it light photometry}~\cite{young2019interferometric}. On the other hand, it has also been shown that the signal measured can also contain structural information about the molecule~\cite{kim2017label}, allowing label-less probing of structural dynamics. While not mutually inconsistent, these contrasting observations 
raise questions regarding how the conformation of a macromolecule affects the strength of the dipole interaction, and whether measurements at the single-molecule level could be used to provide a quantitative understanding associated with the structural dynamics of biological processes such as protein folding~\cite{dill2012protein,dobson2003protein} and enzyme catalysis.

\begin{figure*}[ht!]
\includegraphics[width=\textwidth]{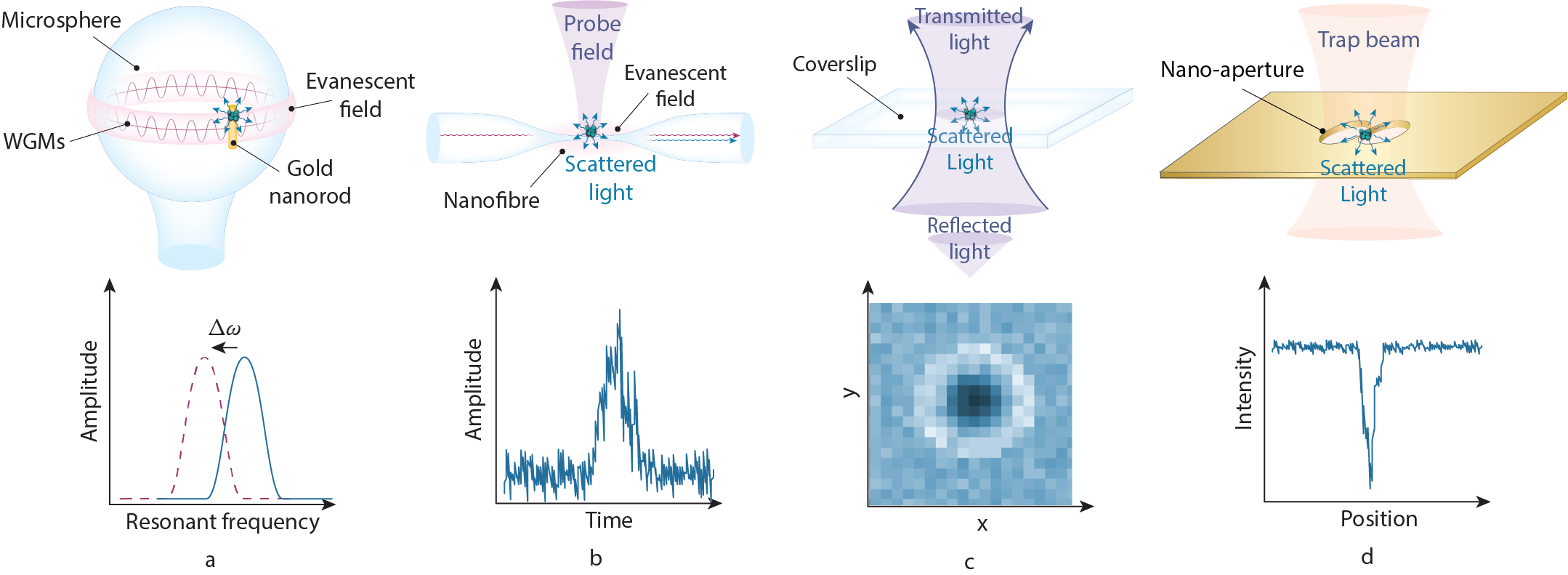}
\begin{flushleft} 
\caption{\label{fig1}} Single-molecule biosensors and their corresponding signals measured. a) Optoplasmonic sensors use a shift in a cavity resonance frequency to detect the presence of a molecule.\cite{baaske2014single} b) Dark-field heterodyne microscopy collects the scattered field from a biomolecule and measures the beat-note of this field with light at a slightly different frequency.\cite{mauranyapin2017evanescent} c) Interferometric scattering microscopy (iSCAT) uses the interference between the light reflected from the glass cover slip and the light scattered from the molecule to construct an image of the molecule\cite{young2018quantitative}. d) Plasmonic tweezers monitor the intensity of light transmitted through a sub-diffraction-limited hole in a plasmonic structure as a biomolecule enters the hole~\cite{Pang2012Optical}.
\end{flushleft}
\end{figure*}

Previous models of dielectric light-matter interactions which have been applied to single-molecule sensors have generally treated the molecule as a homogeneous material~\cite{vollmer2002protein,mauranyapin2017evanescent}, or as a compound structure containing several such homogeneous components~\cite{kim2017label}. These approaches are limited to prediction of large-scale effects such as motion of the molecule out of the sensing region, differential motion of model components, or assembly of multiple molecules into larger structures. They are insensitive to inhomogeneities within the molecule and to changes in its structure. Here, we introduce a method to calculate the dynamic polarizability directly from the atomistic structure of the molecule. The basis of our approach is the interactive dipole model of Applequist {\it et al.}~\cite{applequist1972atom} We combine this model with efficient computational techniques to allow the polarizability of molecules containing more than one hundred thousand atoms to be calculated. We apply our method both to time-domain molecular dynamic simulations, tracking the thermally agitated fluctuations of the polarizability of Bovine Serum Albumin (BSA) as a function of time, and to predict the change in polarizability as the molecular motors ATPase and 26S proteasome change their conformations. We briefly discuss the extent to which differences in polarizability stemming from the changes in conformation could be resolved using state-of-the art single molecule light-scattering measurements. 
Applequist-type approaches are already used to calculate polarizable force fields in molecular dynamics simulations~\cite{Lopes2009Molecular,Halgren2001Polarizable,Duan2016Large-Scale,Xianfeng2005Long,Liu2018Benchmarks,Wang2000How}. These simulations generally take account of both intra-molecule electrostatic forces and dynamic polarization forces. By contrast, our approach deals only with the dynamic polarization in response to high frequency optical excitation ($\sim 10^{12}$ Hz). The aim is to provide a simple method to determine the light scattering properties of macromolecules that is easily accessible to the single molecule sensing community. In line with this aim, we have made the code developed to calculate the dynamic polarizability openly available\cite{Booth2021Code}. We believe that our approach will find applications translating the measured optical signals that arise from the dipole interaction into a quantitative understanding of the structural dynamics of macromolecule, which could be applied to better understand important biological processes such as protein folding and enzyme dynamics.

\section{Theory}

\begin{figure}
\includegraphics[width=\columnwidth]{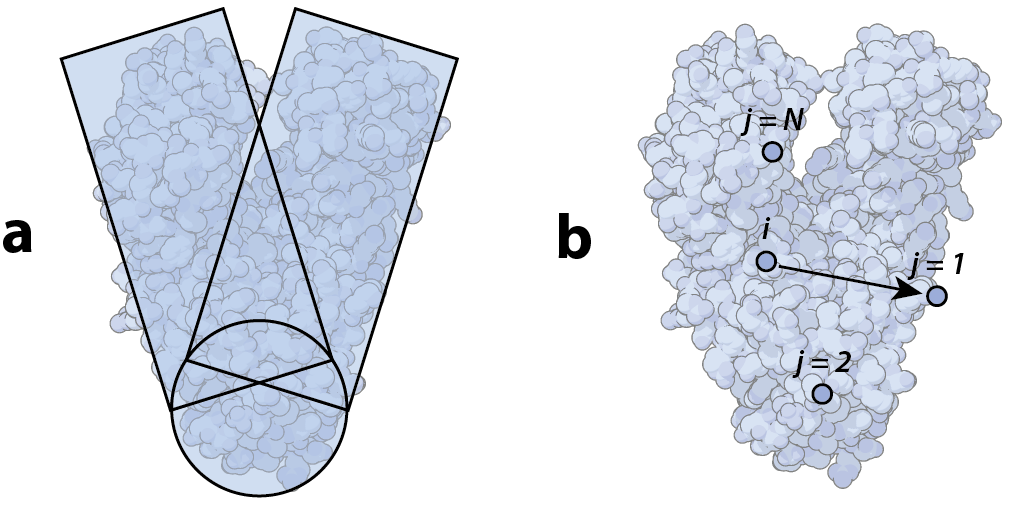}
\caption{\label{fig2} a) Bulk model vs. b) atomistic dielectric model. [Protein Data Bank ID (PDBID): 4YFU], illustrative purposes only.}
\end{figure}

Many of the single-molecule sensors in Fig. \ref{fig1} produce a signal via the dipole interaction between the molecule and light. In this case the size of the signal detected depends on the power scattered from the molecule. The mean power scattered, $\langle P_{sc} \rangle$, is:
\begin{equation}
{ \langle P_{sc} \rangle} = \frac{c\epsilon_0k^4}{96\bm{\pi}{{n}_m}^3} \bm{||\mu_{ex}||}^2,
    \label{scatteredpower}
\end{equation}
where $c$ is the speed of light; $\epsilon_0$ is the vacuum permittivity; $n_m$ is the refractive index of the medium surrounding the molecule; and $k = 2\pi/\lambda$, where $\lambda$ is the wavelength of the incident light and $\bm{\mu_{ex}}$ is the excess dipole moment \cite{jackson_classical_1999}. The excess dipole moment is given by $\bm{\mu_{ex}} = \bm{\mu_{mol}} - \bm{\mu_{m}}$, where $\bm{\mu_{mol}}$ is the molecular dipole moment and $\bm{\mu_{m}}$ is the dipole moment of the surrounding medium displaced by the molecule. It depends on the applied electric field $\bm{E(r)}$ via the relationship
\begin{equation}
\bm{\mu_{ex}} = \bm{\alpha_{ex}} \cdot \bm{E(r)},
    \label{moldipole1}
\end{equation}
where $\bm{\alpha_{ex}}$ is the excess polarizability tensor. Therefore, the power scattered from the molecule depends on its excess polarizability. For the simple case that the molecule is an isotropic sphere, the scattered power is given by the well-known expression \cite{mauranyapin2017evanescent}:
\begin{equation}
\langle P_{sc} \rangle = \frac{\sigma}{8}I_{in} = \frac{k^4\bar{\alpha}_{ex}^2}{48\pi\epsilon_m^2}I_{in},
    \label{scatteredpoweriso}
\end{equation}
where $I_{in}$ is the intensity of the input field in $W/m^2$; $\sigma$ is the scattering cross section of the molecule; $\epsilon_m$ is the permittivity of the surrounding medium; and $\bar{\alpha}_{ex}$ is the average polarizability, defined as $\bar{\alpha}_{ex}$ = $\bar{\alpha}_{av}$ - $\bar{\alpha}_{m}$, where $\bar{\alpha}_{\bm{av}}$ is the average polarizability of the molecule and $\bar{\alpha}_{\bm{m}}$ is the average polarizability of the surrounding medium.

In the field of single-molecule biosensing, the biomolecule is often approximated as a simple dielectric sphere\cite{ mauranyapin2017evanescent} or collection of bulk shapes\cite{kim2017label} (see Fig. \ref{fig2}a). The polarizability and scattered power are then
calculated from bulk properties such as the refractive index of the molecule in solution. However, bulk constants such as refractive index break down at the nanometre scale of single biomolecules. Therefore, alternative methods that account for the atomic structure of molecules are required to accurately calculate their polarizability.

This paper develops an atomistic method based on the interactive dipole model devised by Applequist \textit{et al}. \cite{applequist1972atom} The model assumes that each molecule is a collection of  $N$ atoms, each at fixed spatial coordinates. In the ideal case, this method evaluates both the dipole induced in each atom by an external electric field $\bm{E}$, and the dipole moment induced by the interaction of each atom with every other atom in the molecule (Fig. \ref{fig2}b). The induced dipole moment of the entire molecule ($\bm{\mu_{mol}}$) is the sum of the induced dipole moments of each atom, $\bm{\mu_i}$:
\begin{equation}
\mathbf{\bm{\mu_{mol}}} = \sum_{i=1}^N\bm{\mu_i}.
\label{sumui}
\end{equation}
The induced dipole moment for each atom $i$ is a function of the applied electric field given by 
\begin{equation}
    \bm{\mu_i}(\bm{E}(\bm{r})) = {\alpha}_i \left[ \bm{E}(\bm{ r_i}) - \sum_{j\neq i}^N {\mathbf{T}_{ij} \bm{\mu}_j} \right],
    \label{induceddm}
\end{equation}
where $\alpha_i$ is the atomistic polarizability of atom $i$, $\bm r_i = \{x_i, y_i, z_i\}$ is the position of the atom $i$, and $\bm{T_{ij}}$ is the dipole field tensor given by Ref. \onlinecite{thole1981molecular}:
\begin{equation}
    \mathbf{T}_{ij} = 
\begin{cases}
    \hspace{0.525cm} \frac{1}{\Delta r^3} \mathbf{I} - \frac{3}{\Delta r^5}
    \begin{bmatrix}
    \Delta x^2 & \Delta x \Delta y & \Delta x \Delta z \\
    \Delta x \Delta y & \Delta y^2 & \Delta y \Delta z \\
    \Delta x \Delta z & \Delta y \Delta z & \Delta z^2
    \end{bmatrix}\vspace{0.5cm},& \text{if } \Delta r>s\\

    \frac{4v^3-3v^4}{\Delta r^3} \mathbf{I} - \frac{3v^4}{\Delta r^5}
    \begin{bmatrix}
    \Delta x^2 & \Delta x \Delta y & \Delta x \Delta z \\
    \Delta x \Delta y & \Delta y^2 & \Delta y \Delta z \\
    \Delta x \Delta z & \Delta y \Delta z & \Delta z^2\end{bmatrix}, & \text{if } \Delta r<s.
\end{cases}
    \hspace{1.3cm} 
    \label{dipolefieldtensor}
\end{equation}
\vspace{0.5cm}
Here $\mathbf{I}$ is the identity matrix, $\{\Delta x,  \Delta y,  \Delta z\} = \{x_i-x_j, y_i-y_j, z_i-z_j\}$, $\Delta r \smash{= \sqrt{\Delta x^2 + \Delta y^2 + \Delta z^2}}$ is the distance between atoms $i$ and $j$, $v=\Delta r/s$, and $s = 1.662(\alpha_i\alpha_j)^{1/6}$. It includes a filter function to ‘smear’ out the dipoles to avoid a polarization catastrophe developed by Thole {\it el al.}~\onlinecite{thole1981molecular}, where the magnitude of the dipole interaction can become infinite under certain conditions\cite{Cieplak2009Polarization}.

The polarizability vector $\bm{\alpha_k}$ defines the response of a molecule to a given electric field, where the subscript $\bm{{k}}$ denotes the direction of the electric field. Given a uniform electric field across the molecule such that $\bm{E_k}(\bm{r_i})=\bm{E_k}$ at all atoms, the polarizability vector can be calculated from the induced dipole moment in Eq.~\ref{amol,k} as
\begin{equation}
\noindent\mathbf{\bm{\alpha_{{k}}}} = \sum_{i=1}^N\frac{\bm{\mu_i}(\bm{E_{{k}}})}{\bm{||E_{{k}}||}}.
\label{amol,k}
\end{equation}
The polarizability tensor of the molecule is then given by the combination of the polarizability vectors for three orthogonal electric field directions ${\bm{x}}$, ${\bm{y}}$ and, ${\bm{z}}$:

\begin{equation}
    \bm{\alpha_{mol}} =
    \begin{bmatrix}
    
    \begin{bmatrix}
    \vspace{4mm}\\
\bm{\alpha_{{\bm{x}}}}\\
\vspace{4mm}
 \end{bmatrix}    
 
 \begin{bmatrix}
    \vspace{4mm}\\
\bm{\alpha_{{\bm{y}}}}\\
\vspace{4mm}
 \end{bmatrix}
 
  \begin{bmatrix}
    \vspace{4mm}\\
\bm{\alpha_{{\bm{z}}}}\\
\vspace{4mm}
 \end{bmatrix}

    \end{bmatrix}.
    \label{amol}
\end{equation}

\noindent The average polarizability is\cite{applequist1972atom} 
\begin{equation}
\bar{\alpha}_{av} = \frac{\lambda_1 + \lambda_2 + \lambda_3}{3},
\label{averagepol}
\end{equation}
where $\lambda_1, \lambda_2$ and $\lambda_3$ are the eigenvalues of $\bm{\alpha_{mol}}$.

\section{Computational methods}

To calculate the polarizability from the spatial coordinates of each atom in the molecule, we evaluate the induced dipole moments of every atom using Eq.~\ref{induceddm}, which can be arranged as:
\begin{equation}
    \widetilde{\mathbf{A}} \widetilde{\mathbf{\mu}} = \widetilde{\mathbf{E}},
    \label{matrixeq}
\end{equation}
where
\begin{equation}
     \widetilde{\mathbf{A}}=
    \begin{bmatrix}
    \bm{\alpha}_1^{-1} & \mathbf{T}_{12} & \dots & \mathbf{T}_{1N} \\
    \mathbf{T}_{21} & \bm{\alpha}_2^{-1} & \dots & \mathbf{T}_{2N} \\
    \vdots & \vdots & \ddots & \vdots \\
    \mathbf{T}_{N1} & \mathbf{T}_{N2} & \dots & \bm{\alpha_N}^{-1}
    \end{bmatrix}
     \label{A}
\end{equation}
is a $3N\times3N$ matrix containing the polarizability tensors $\mathbf{T}_{ij}$ of each pair of atoms on its off-diagonals, and on its diagonals the atomic polarizability matrices for each atom $i$, given by
\begin{equation}
   \bm{\alpha}_i^{-1} = 
    \begin{bmatrix}
   \alpha_i^{-1} & 0 & 0\\
  0 &  \alpha_i^{-1}  & 0\\
0 & 0 & \alpha_i^{-1}
    \end{bmatrix}.
    \label{pol_diag_matrix}
\end{equation}
 $\widetilde{\mathbf{E}}$ is a $1\times3N$ vector containing the electric field at each atom and $\widetilde{\mathbf{\mu}}$ is a $1\times3N$ vector containing the induced dipole moment of each atom; i.e. 
 \begin{equation}
    \widetilde{\mathbf{\mu}}=
    \begin{bmatrix}
    \bm{\mu_1}\\
    \bm{\mu_2}\\
    \vdots\\
    \bm{\mu_N}
    \end{bmatrix}, \hspace{5mm}
       \widetilde{\mathbf{E}}=
    \begin{bmatrix}
   \bm{E_1}\\
   \bm{E_2}\\
   \vdots\\
   \bm{E_N}
    \end{bmatrix},
    \label{Au=E}
\end{equation}
where $\bm{E_i} = [E_{i,x}, E_{i,y}, E_{i,z} ]^{\rm T}$ and $\bm{\mu_i} = [\mu_{i,x}, \mu_{i,y}, \mu_{i,z} ]^{\rm T}$. As can be seen from Eq.~\ref{matrixeq}, $\widetilde{\mathbf{\mu}}$ can be found from the inverse of $\widetilde{\mathbf{A}}$ as $\widetilde{\mathbf{\mu}} = \widetilde{\mathbf{A}}^{-1} \widetilde{\mathbf{E}}$. Assuming a uniform electric field, the polarizability vector of the molecule can then be calculated using Eq.~\ref{amol,k}, and from this the polarizability tensor  $\bm{\alpha_{mol}}$ and average polarizability $\bar\alpha_{av}$ can be calculated using Eqs.~\ref{amol}~ and~\ref{averagepol}, respectively. 

\subsection{Testing with small molecules}

\begin{table}[h]
\begin{ruledtabular}    
\caption{Atomic polarizabilities used in calculations}
\begin{tabular}{cc}
Atom & Polarizability ($\text{\AA}^3$) \\ \hline
H & 0.514\footnotemark[1] \\
C & 1.405\footnotemark[1] \\
N & 1.105\footnotemark[1] \\
O & 0.862\footnotemark[1] \\
S & 2.900\footnotemark[2] \\
P & 3.630\footnotemark[2] \\
\end{tabular}
\footnotetext[1]{Values from Ref.~\onlinecite{thole1981molecular}, $\lambda$ = 589.3 nm}
\footnotetext[2]{Values from Ref.~\onlinecite{miller1978atomic}}
\end{ruledtabular}
\end{table}

To verify the accuracy of the model, it was tested on small molecules with known experimental bulk average polarizabilities.  The values for the atomic polarizability $\alpha_i$ used in this paper are displayed in Table I, and the $x,y,$ and $z$ co-ordinates of each atom in a given molecule were obtained from Ref.~\onlinecite{Chemical}. Henceforth, we assume a uniform electric field $\bm{E_k}$ across the molecule.  However, the induced dipole moment, and therefore the scattered power, could be calculated for any applied field, such as the highly nonuniform fields typical of plasmonic biosensors\cite{Shopova2011Plasmonic}$^,$\cite{Swaim2011Detection}. 

Table II compares the average molecular polarizability calculated here (column C) to those reported by Applequist \textit{et al.}\cite{applequist1972atom} (column A), Thole\cite{thole1981molecular} (column T), and experimental values\cite{applequist1972atom} (column E). Applequist \textit{et al.} reports values for average polarizability that differ from calculated values by up to 22~\% (for H$_2$O). We attribute these discrepancies to the polarization catastrophe discussed earlier. Indeed, we found this catastrophe to become increasingly problematic as the molecular size became larger, so that it was not possible to obtain accurate results without Thole's correction. Our model, as described so far, is identical to the one used by Thole, and yielded very similar results (within 1~\%) with minor differences attributed to the use of slightly different molecular coordinates. 
As shown in the table, both sets of results agree well with the experimental average polarizability, with a maximum deviation of 4~\%.

\begin{table}[h]
\begin{ruledtabular}
\caption{Experimental vs. calculated polarizabilities for various small molecules: Experimental (E), Applequist \textit{et al.} (A), Thole (T), Calculated (C). $\lambda$ = 589.3 nm}
\centering
\begin{tabular}{lcccc}
\multicolumn{1}{l}{} & \multicolumn{4}{c}{Average polarizability{($\text{\AA}^3$)}}   \\
Compound & E\footnotemark[1] & A\footnotemark[1] & T\footnotemark[2] & C \\ \hline
Methane ($\text{CH}_4$) & 2.62 & 2.58 & 2.55 & 2.55 \\
Ethane ($\text{C}_2\text{H}_6$) & 4.48 & 4.47 & 4.46 & 4.43 \\
Propane ($\text{C}_3\text{H}_8$) & 6.38 & 6.58 & 6.29 & 6.32 \\
Cyclohexane ($\text{C}_6\text{H}_{12}$) & 11.00 & 10.95 & 10.95 & 11.02 \\
Formaldehyde ($\text{CH}_2\text{O}$) & 2.45 & 2.46 & 2.54 & 2.51 \\
Dimethyl ether ($\text{C}_2\text{H}_6\text{O}$) & 5.24 & 5.22 & 5.24 & 5.26 \\
Acetone (($\text{CH}_3)_2\text{CO}$) & 6.39 & 6.44 & 6.32 & 6.34 \\
Water ($\text{H}_2\text{O}$) & 1.49 & 1.12 & 1.44 & 1.43 \\
\end{tabular}
\end{ruledtabular}
\footnotetext[1]{Values from Ref.~\onlinecite{thole1981molecular}}
\footnotetext[2]{Values from Ref.~\onlinecite{applequist1972atom}}
\end{table}

\subsection{Scaling up to larger molecules}

Motor molecules and other proteins are much larger than the molecules computed in Section 3.1 (for example, the motor molecule ATPase has $N$ = 39,000 atoms). As there are 9$N^2$ elements in $\widetilde{\mathbf{A}}$, computer memory constraints fast become a problem when trying to compile $\widetilde{\mathbf{A}}$ for larger molecules. In our case, this has prevented application of the method for $N$ > 17,000. 

The matrix $\widetilde{\mathbf{A}}$ contains the polarizability tensor of each atom as well as the dipole field tensors for each dipole formed by two atoms (Eq.~\ref{A}). The magnitude of the dipole field tensor scales as $1/\Delta r^5$. Therefore, as the distance between pairs of atoms increases their contribution to the overall polarizability falls rapidly and can be neglected (set to zero). This leads to a sparse matrix and a dramatic reduction in the total number of matrix elements that need to be stored. To test this, we introduce a threshold radius $\Delta r_t$, and set $\mathbf{T}_{ij}$ = 0 for all elements in $\widetilde{\mathbf{A}}$ for which $\Delta r > \Delta r_t$. 

 The polarizability determined for a single protein with and without a threshold radius is compared in Fig. \ref{fig7}a. Bovine Serum Albumin (BSA) is used for this comparison. BSA is commonly used in label-free single molecule experiments\cite{vollmer2002protein}$^,$\cite{ mauranyapin2017evanescent}. It contains 5.944 atoms, which is below the size at which memory errors become a problem ($N$ = 17,000). 
At threshold radii below 2 $\text{\AA}$, the thresholded calculation overestimates the polarizability of BSA by approximately 10~\%. 
This is because all atoms are separated by greater distances than $\text{\AA}$, and so all atom-atom dipoles are excluded from the molecular dipole. The resulting molecular polarizability is then just the sum of the atomic polarizabilities of each atom ($\alpha_0$). At thresholds close to but above 2 $\text{\AA}$, the polarizability is underestimated by a similar margin.  This indicates that including only interactions from closely separated atoms causes a decrease in the estimated polarizability. As atomic interactions with greater separations are included, the polarizability asymptotes towards the value obtained with no threshold radius applied. For instance, for a threshold radius of $\Delta r_t$ = 30 $\text{\AA}$, the polarizability is within 99~\% of the threshold-less value. Henceforth, we use this 30 $\text{\AA}$ threshold radius in our calculations. We found that this significantly reduced memory requirements, allowing calculations for molecules as large as $N$ = 110,000. 

We note that the calculations in Fig. \ref{fig7}a shed some light on when it may be appropriate to use a bulk model or a model based on a collection of sub-domains to calculate the polarizability (e.g. see Ref. \onlinecite{kim2017label}). If the effective radii of each sub-domain is greater than one nanometer, then it can be expected that interactions between the sub-domains will not affect the polarizability significantly. For smaller, closely connected sub-domains, the interactions will be significant, and they should be explicitly considered in any model used to determine the polarizability. For instance, an aggregate of several BSA molecules. Since each molecule is roughly 3~nm in radius, after performing an atomistic calculation of the polarizability of one BSA molecule, the polarizability of the aggregate could be expected to be estimated by treating it as a compound system of non-interacting parts, each with the same polarizability.

\begin{figure}[ht!]
 \begin{tabular}{@{}c@{}}
\includegraphics[width=\columnwidth]{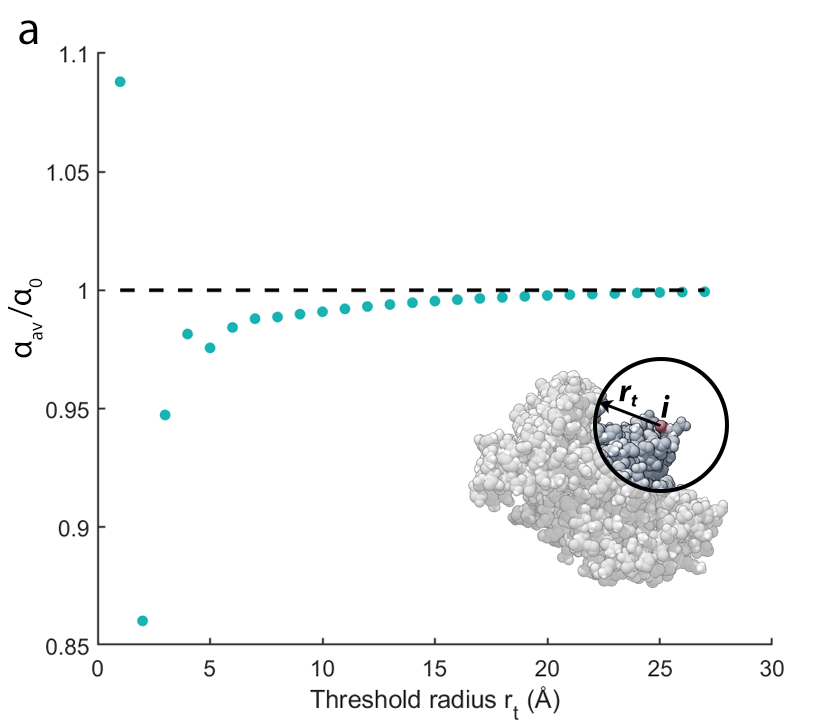}
\end{tabular}
  \vspace{\floatsep}

 \begin{tabular}{@{}c@{}}
\includegraphics[width=\columnwidth]{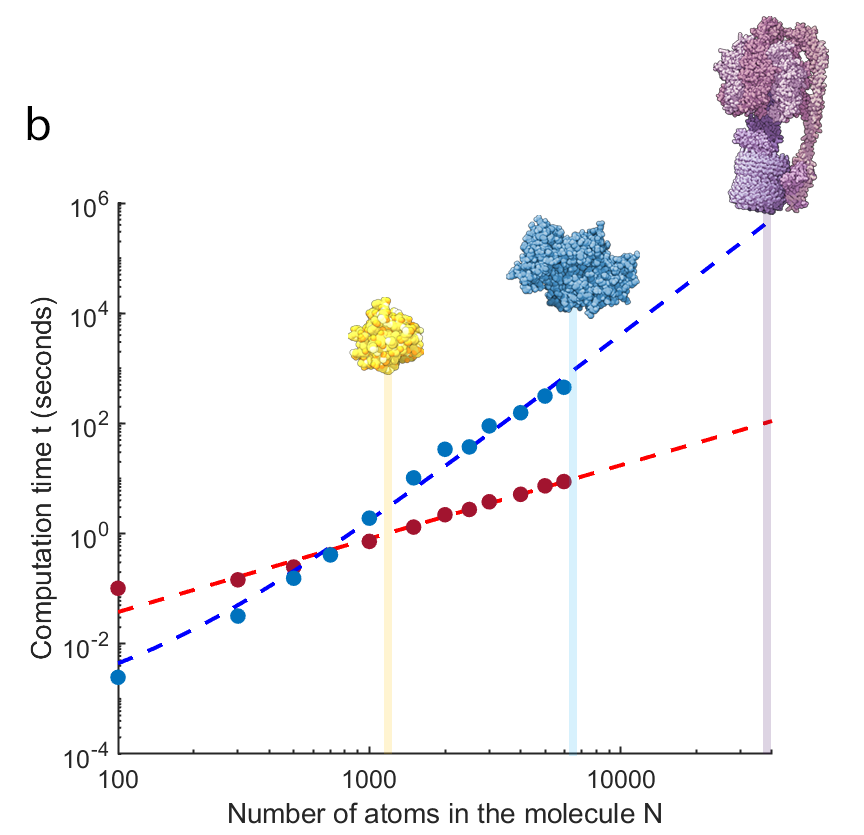}
\end{tabular}
\caption{\label{fig7} a) Threshold radius vs. average molecular polarizability ($\alpha_{av}$). $\alpha_0$ is the average polarizability of BSA [PDBID: 3V03]\cite{Majorek:2012} with no threshold radius applied. b) Comparison of the time taken to calculate $\widetilde{\mathbf{\mu}}$ by inverting $\widetilde{\mathbf{A}}$ (in blue) and by using an iterative solver (in red). Sizes of various proteins (Photoactive yellow protein [PDBID: 3VE4], BSA, and ATPase [PDBID: 6FKI]).}
\end{figure}

Eq.~\ref{matrixeq} can be solved for $\widetilde{\mathbf{\mu}}$ by inverting $\widetilde{\mathbf{A}}$. However, even
with a sparse matrix we found that the time required to
perform this operation increased rapidly with molecule
size, making the cost of this approach prohibitive even for
relatively small proteins. This is shown quantitatively in Fig.
3b. The computational time scales as $ N^{3.5}$ for molecules with
more than 500 atoms. To address this limitation, the use of
an iterative solver, the MATLAB minimum residual
method (minres), was examined. While this solver iterates towards a solution to the matrix inverse problem, rather than solving directly, we find that it offers sufficient accuracy with greatly reduced time overhead when dealing with large molecules. For small molecules ($N$ < 700), directly inverting $\widetilde{\mathbf{A}}$ is faster. However, the iterative method scales as $ N^{1.3}$ (see Fig. \ref{fig7}b) allowing a dramatic speed up for large molecules. For instance, it was possible to calculate $\widetilde{\mathbf{\mu}}$ for ATPase ($N$ = 39,000) in 106 seconds, compared to 130 hours using a direct matrix inverse.

A caveat for the iterative method is that it requires a tolerance for accuracy to be set before the calculation. The solver will continue to approximate the result over a given number of iterations until the residual of the result is below this tolerance level. Increasing the tolerance can decrease run time but it can also decrease accuracy of the result. In this paper, we use a tolerance of 10$^{-6}$, which is the default tolerance for the minres function in MATLAB.

\subsection{Testing with bulk materials}

Combining the sparse matrix and iterative solver discussed in section 3.2, we were able to extend the polarizability calculations to relatively large molecules, and to test the method with bulk materials that have well-known refractive indices. Polarizability is related to the refractive index via the Lorentz-Lorenz relation:
\begin{equation}
\frac{n^2-1}{n^2+2} = \frac{4\pi}{3}\rho \bar{\alpha}_{av},
\label{LL}
\end{equation}
where $n$ is the refractive index and $\rho$ is the atomic density per cubic metre\cite{LorentzLorenz}. 

\begin{table}[h]
\begin{ruledtabular}
\caption{Polarizability of bulk materials}
\centering
\begin{tabular}{lccc}

Material & $\bar{\alpha}_{av}$($\text{\AA}^3$)  & Calculated $n$ & Experimental $n$ \\ \hline
Diamond & 4030 & 2.458 & 2.417\footnotemark[1] \\
H$_2$O & 3944 & 1.3299 & 1.3325 \\

\end{tabular}
\end{ruledtabular}
\label{tab3}
\footnotetext[1]{Values from Ref.~\onlinecite{Phillip1964Kramers-Kronig}. Experimental refractive indices are at a wavelength of $\lambda = 589.3~nm$}
\end{table}

The polarizability of two bulk materials, diamond and water, was calculated. For diamond, we used a crystal lattice constant of $a = 3.567~\text{\AA}$ to generate a set of atomic coordinates\cite{Duke2002Frontiers}. For water, we used a snapshot taken from a molecular dynamics simulation of liquid water at 273 K at atmospheric pressure. The water was described using the TIP5P model\cite{Abraham:2015aa}. Table III shows the refractive indices calculated based on these structural models using our approach versus experimental values. The calculated refractive indices for diamond and water are within 2~\%  and 0.2~\% of the experimental values, respectively. This close agreement provides some confidence in the ability of the model to accurately calculate the molecular polarizability of biomolecules.

\subsection{Code availability}

The code developed for this paper, using both sparse matrices and iterative matrix inversion, is available at Ref.~\cite{Booth2021Code}.

\section{Results}

\subsection{Polarizability of Bovine Serum Albumin} 

As a first application of our method to larger molecules, we considered the case of a single BSA molecule [PDBID: 3V03]\cite{Majorek:2012}. Our approach yields an average molecular polarizability $\bar{\alpha}_{av}$ of $5,860~\text{\AA}^3$, averaged over a 54 ns simulation with 1080 data points. Two replicate molecular dynamics simulations yielded the same value for $\bar{\alpha}_{av}$ within 0.2~\%. These replicate simulations used the same Protein Data Bank (PDB) file for BSA [PDBID: 3V03]\cite{Majorek:2012} but had different starting velocities generated from a Maxwell-Boltzmann distribution at 300 K.

For comparison, Vollmer {\it et al.} use a bulk model to calculate an excess polarizability for BSA of $3,850~\text{\AA}^3$~\cite{vollmer2002protein} which, accounting for the polarizability of the displaced water, corresponds to a molecular polarizability of around $7,800~\text{\AA}^3$. This is reasonably close to the result of our calculation. The good agreement of our model with known results for bulk materials is indicative that the discrepancy may arise from the use of a bulk model in Ref.~\onlinecite{vollmer2002protein}.

\subsection{Conformational changes due to thermal fluctuations}

\begin{figure}[ht!]
\includegraphics[width=\columnwidth]{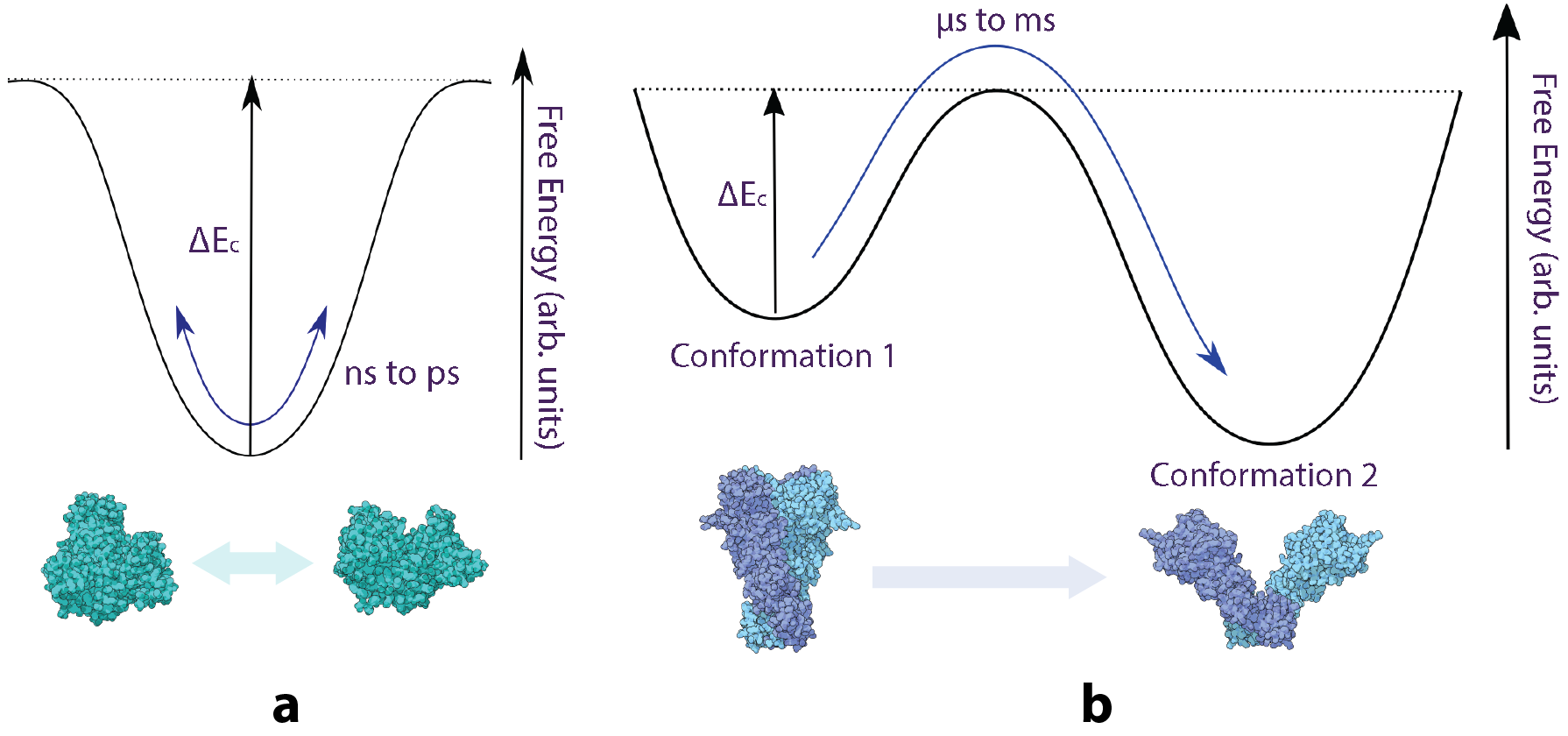}
\caption{\label{fig3}
a) Fluctuations in polarizability on a picosecond-nanosecond time scale are typically due to thermal fluctuations of a single molecular conformation \cite{henzler2007dynamic}, [PDBID: 3V03]\cite{Majorek:2012}, illustrative purposes only. b) Free energy diagram of a conformational change of a molecule, which typically occurs over a time scale of microseconds to milliseconds \cite{henzler2007dynamic}, [PDBID: 2CG9, 2IOQ], illustrative purposes only}
\end{figure}

So far, we have treated each biomolecule as a rigid structure, that is, a collection of atoms with fixed coordinates. However, in reality proteins are highly flexible. They reside in multiple local minima at room temperature and fluctuate around these local minima on a wide variety of timescales. This is illustrated diagrammatically in Fig.~\ref{fig3}. Fig.~\ref{fig3}a represents motion around a local minima while Fig.~\ref{fig3}b represents two alternative quasi-equilibrium conformational states separated by an energy barrier. An attractive aspect of the methodology we have developed is that it opens the possibility to probe structural transitions in single molecules on the timescales accessible to molecular dynamics simulation techniques. In principle this makes it possible to directly relate temporal fluctuations in the polarizability measured experimentally to changes in underlying molecular configuration as probed using atomistic simulations. As an illustration of this potential we have compared temporal measurements of the polarizability of a single molecule of BSA freely rotating in aqueous solution at 273 K to results from a series of 54~ns molecular dynamics simulations of BSA described using the GROMOS 54A7 force field\cite{Schmid:2011, Spoel:2005, Poger:2010} in TIP5P water\cite{Abraham:2015aa}. See supplementary for more information.

 Figure \ref{fig5}a plots the change in the average excess polarizability $\bar{\alpha}_{ex}$ of 
 BSA at 50~ps intervals from a typical simulation. The average polarizability varies by around 2.5~\% over the simulation. The two replicate molecular dynamics simulations for BSA yielded similar variation of the average polarizability over the same time period (3~\% and 2.5~\%). Note, the average polarizability is unaffected by the orientation of the molecule. It is only sensitive to variations in conformation. In this case the conformations sampled correspond to a system sampling a local minmima or in a quasi-steady state (illustrated in Fig.~\ref{fig3}a).

Under illumination from a spatially uniform light field with optical polarisation $k$, the scattered optical power from the molecule (Eq.~\ref{scatteredpower}) is determined by the magnitude of the excess polarizability vector, defined as $||\bm{\alpha_{\bm{ex,k}}}|| = ||\bm{\alpha_{\bm{k}}}|| - \bar{\alpha}_{\bm{m}}$. Fig.~\ref{fig5}b plots this magnitude for BSA for three orthogonal optical polarizations ($k = x,y,$ and $z$). From this value, we can directly infer the scattered power for each conformation (right axis). 
The polarizability vector (as defined in Eq.~\ref{amol,k}) depends on the direction of the electric field relative to the molecule and is sensitive to the rotation of the molecule as well as to any deformations.  With an applied electric field in the $x$-direction, the magnitude of the excess polarizability vector fluctuated by around 27~\% over the course of the 54 ns simulation. For electric fields applied in the $y$- and $z$-direction, it changed by 19~\% and 16~\% respectively. That the change in ||$\bm{\alpha_{\bm{ex,k}}}$|| is up to 10 times greater than change in average excess polarizability indicates that rigid rotation of the BSA molecule dominates over thermally driven deformations.

Fluctuations in light scattering due to thermally-driven changes in the excess polarizability vector are a potential source of error in single-molecule sensing techniques such as mass photometry~\cite{young2018quantitative}. Our BSA simulations allow us to roughly estimate the magnitude of such errors. Taking, for example, an applied electric field in the $x$-direction, and assuming that the fluctuations in the  excess polarizability are Gaussian, we find that the standard-error in the excess polarizability vector is around 0.13~\%. We conclude, therefore, that thermally-driven fluctuations in the excess polarizability can be expected to be a negligible source of noise, even for a 50~ns measurement. We note that the standard-error typically scales as the inverse-square-root of measurement time, and should therefore be  negligible in realistic scenarios, where the measurement duration is on the order of milliseconds~\cite{mauranyapin2017evanescent}. This is consistent with observations in Ref.~\onlinecite{young2018quantitative}.

\begin{figure*}[ht]
\includegraphics[width=\textwidth]{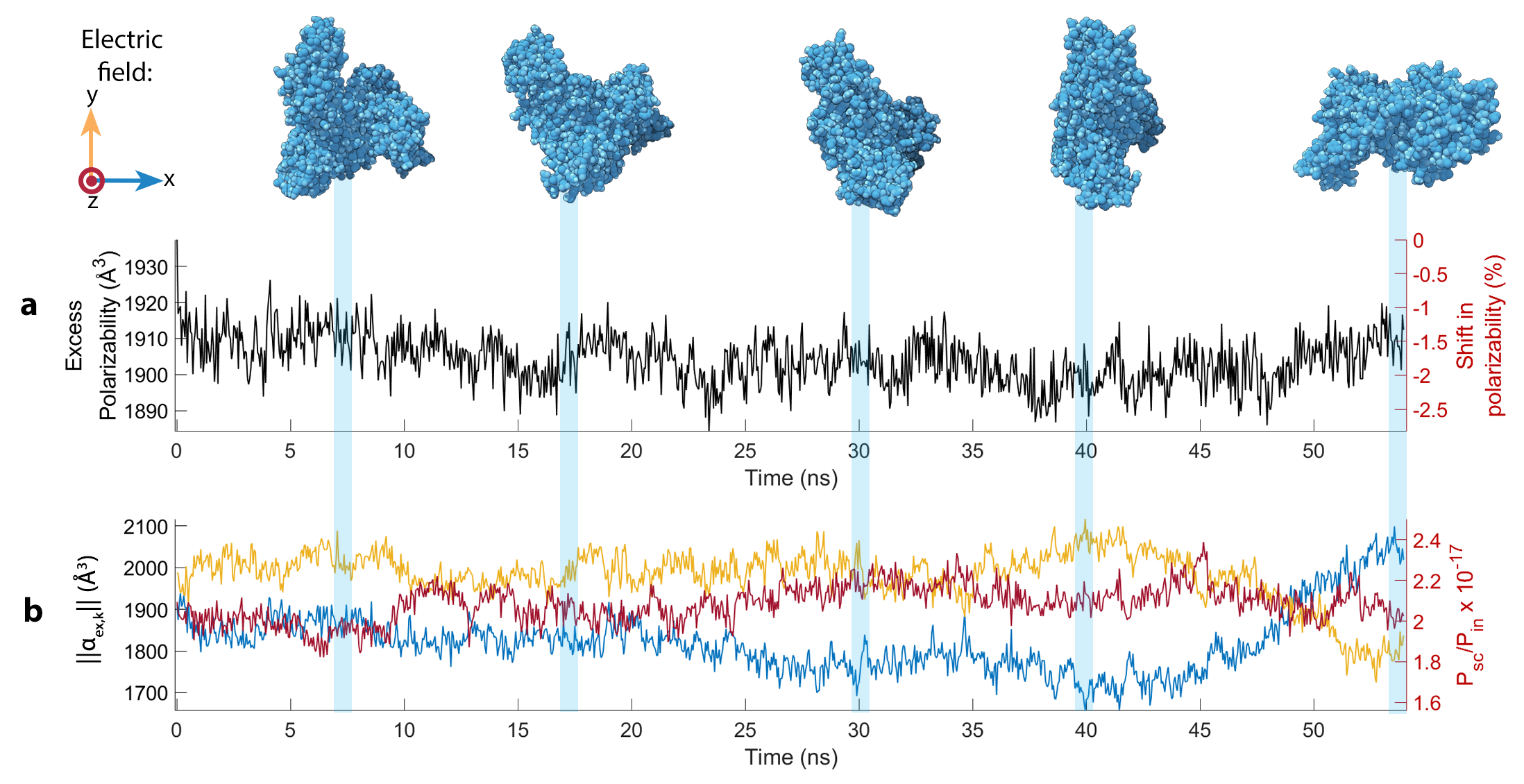}
\caption{\label{fig5}a) Average excess polarizability fluctuations of BSA over 54 nanoseconds. b) Magnitude of the polarization vector of BSA with polarized input light (in red, yellow and blue), with arrows indicating the optical electric field direction. [PDBID: 3V03]\cite{Majorek:2012}} 
\end{figure*}

\subsection{Distinguishing between different quasi-equilibrium conformational states}

The measurements and simulations in Fig. \ref{fig5} represent a system that is close to a single quasi-equilibrium state. Transitions between alternative quasi-equilibrium states (as illustrated in Fig. \ref{fig3}b\cite{henzler2007dynamic}) are key to the function of many proteins. Coordinates for thermally accessible alternative conformations of a wide range of proteins are available from the Protein Data Bank\cite{PDB}. We have used our methodology to assess whether transitions between these alternative conformations might be detected experimentally based on predicted changes in polarizability. For this, two molecules have been considered: Chloroplast F1F0 ATPase [PDBID: 6FKF, 6FKH, and 6FKI]\cite{Hahn2018Structure} (Fig. \ref{conf_changes}a) and 26S Proteasome [PDBID: 6FVT, 6FVU, 6FVV, 6FVW, 6FVX, and 6FVY]\cite{Eisele2018Expanded} (Fig. \ref{conf_changes}b).

Chloroplast $\text{F}_1\text{F}_0$ ATPase is a rotary enzyme complex that uses a proton gradient across the thylakoid membrane to drive the production of ATP. It has a mass of 597 kDa, and consists of roughly 39,000 atoms. The enzyme complex is believed to adopt three distinct conformational states during each rotation. The spatial coordinates of the chloroplast $\text{F}_1\text{F}_0$ ATPase atoms in each of these three conformational states have been inferred from cryo-EM data\cite{Hahn2018Structure} and are illustrated in Fig. \ref{conf_changes}. The rotor consists of c subunits (coloured in white-blue gradient), and central stalk subunits ($\gamma$ in purple and $\epsilon$ in lilac). Between conformation 1 and 2 the rotor turns 112\degree, and between conformation 2 and 3 it rotates 103\degree. The remaining 145\degree~ is the angle between the conformation 3 and 1\cite{Hahn2018Structure}. The rotor of labelled ATPase completes one full rotation in 2.2 ms \cite{nakanishi-matsui2006Stochastic}.

\begin{figure*}[ht]
\includegraphics[width=\textwidth]{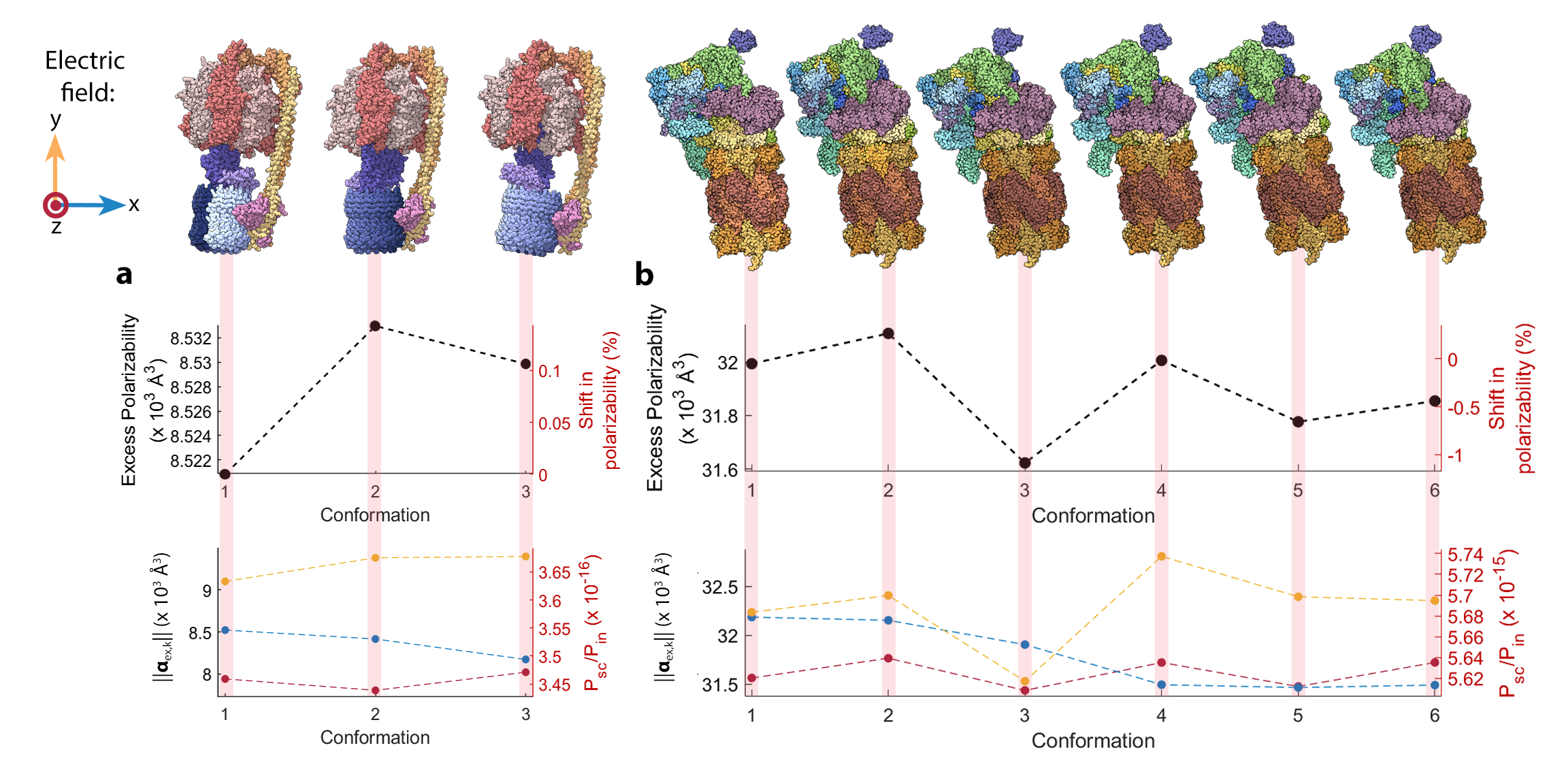}
\caption{\label{conf_changes} Average excess polarizability \textit{(top)} and the magnitude of the polarizability vector for three orthogonal electric field vectors ($k = x,y,$ and $z$) \textit{(bottom)} for different conformations of a) chloroplast $\text{F}_1\text{F}_0$ ATP synthase [PDBID: 6FKF, 6FKH, and 6FKI] and b) 26S proteasome [PDBID: 6FVT, 6FVU, 6FVV, 6FVW, 6FVX, and 6FVY].}
\end{figure*}

Fig. \ref{conf_changes}a{ (\it top)} shows a plot of the average excess polarizability of chloroplast $\text{F}_1\text{F}_0$ ATPase in each of its three conformational states. The average excess polarizability increases by 0.14~\% from conformation 1 to 2,  and decreases by 0.04~\% from conformation 2 to 3. The average polarizability then decreases by 0.1~\% between conformation 3 and 1. Since the average excess polarizability is insensitive to the orientation of the molecule, these differences stem from changes in conformation.

The magnitude of the polarizability vector ||$\bm{\alpha_{\bm{ex,k}}}$|| of each ATPase conformation is plotted in Fig. \ref{conf_changes}a{ (\it bottom)} under three applied optical electric field directions ($k = x,y,$ and $z$). The protein has been orientated such that the $x$, $y$ and $z$ axes correspond to those in the protein's PDB coordinate file. With the electric field applied along the \textit{x}-axis, the polarizability decreases by 1.2~\% between conformations 1 and 2, and decreases a further 2.9~\% between conformation 2 and 3. It then increases by 4.1~\% between conformation 3 and 1. For an electric field along the \textit{y}-axis, the magnitude of the polarizability  increases by 3.1~\% between conformation 1 and 2, and a further 0.2~\% between conformation 2 and 3, decreasing by 3.3~\% between conformation 3 and 1. For a \textit{z}-axis electric field, the polarizability magnitude first decreases by 1.7~\% between conformation 1 and 2, then increases by 2.7~\% between conformation 2, and decreases by 1~\% between conformation 3 and 1.

These changes in polarization vector magnitude likely arise due to large scale reorganisation that occurs between conformations. The primary change is the rotation of the central stalk. Note, the $\epsilon$ subunit (coloured lilac in Fig.~\ref{conf_changes}a) moves around the outside of the central stalk as it rotates. The ATPase molecule also tilts at different angles relative to the $y$-axis in each conformational state. For example, conformations 2 and 3 are tilted by 10.2\degree~ and 11.5\degree~ relative to conformation 1, respectively.

In the case of chloroplast $\text{F}_1\text{F}_0$ ATPase, the changes in the polarizability vector between conformations are about an order of magnitude greater than the changes in average polarizability. For instance, the average polarizability increases by 0.14~\% between conformation 1 and 2, compared to 3.2~\% for the magnitude of the \textit{y}-axis polarizability vector.  Changes in average polarizability are likely to predominantly result from a change in the average separation of atoms, and therefore to correspond roughly to a change in the density of the molecule.

As a second example molecule that undergoes conformational changes, we consider  26S proteasome (Fig. \ref{conf_changes}b). 
Based on cryo-EM data for 26S Proteasome (mass 1,583~kDa, approximately 110,000 atoms) a series of six conformational states or structural classes have been identified (s1-s6)~\cite{Eisele2018Expanded}. Conformational states are dominated by the six ATPase subunits, which undergo concerted cycles of ATP binding, hydrolysis and release\cite{Eisele2018Expanded}, together with the gate or cap which is closed in conformations s1,s2, and s3 and open in conformations s4,s5, and s6\cite{Eisele2018Expanded}.

Fig. \ref{conf_changes}b (\textit{top}) plots the average polarizability predicted for the six major conformational states of 26S proteasome. The average polarizability of proteasome varies between 31,600 $\text{\AA}$ (conformation 3) and 32,100 $\text{\AA}$ (conformation 2). This is a change of approximately 1~\% between conformations, ten times greater than that of ATPase. 

Fig. \ref{conf_changes}b (\textit{bottom}) shows plots of the magnitude of the polarizability vector when an $x$, $y$ or $z$ aligned optical field is applied for the six conformational states of 26S proteasome examined. For an electric field aligned along the \textit{x}-direction, the magnitude of the polarizability vector is 32,000 $\text{\AA}^3$ or above for s1, s2, and s3 (gate closed), but decreases to 31,500 $\text{\AA}^3$ for s4, s5, and s6 (gate open). This suggests that the opening and closing of the gate could potentially be resolved from changes in the scattering cross-section of the molecule. This would require that the 26S proteasome is bound to a surface to prevent free rotation that would result in an averagiung of the polarizability vectors for the three optical field alignments. With a \textit{y}-axis electric field, the magnitude of the polarizability vector dips notably for conformation s3, a dip that is also observed in the average polarizability. When applying a \textit{z}-axis electric field, the magnitude of the polarizability vector changes the least, with maxima in the s2, s4 and s6 states.

26S Proteasome is composed of thirty-three unique proteins that reorganise between conformations. Thus it is difficult to infer what exact structural differences are causing the changes in polarizability predicted by our model. Unlike ATPase, however, the changes in average polarizability between 26S proteasome conformations are on the same order of magnitude as the changes in the magnitude of the polarizability vector. This suggests that changes in inter-atom separation play a significant role in the differences in polarization between conformations.

\subsection{Detectability via optical scattering}

Molecular conformational changes have been successfully detected using optoplasmonic sensors~\cite{kim2017label}. In this approach the high spatial gradients that exist in the electric field of a plasmonic resonator are used. Motion  shifts a significant fraction of the molecule out of the plasmonic hot spot, resulting in a reduction in scattering.
It is interesting to ask whether the polarizability changes predicted here could also provide a means to observe changes in molecular conformation. That is, whether the changes in optical scattering due to alterations in molecular conformation could be resolvable using state-of-the-art single-molecule biosensors such as those in Ref.~\onlinecite{mauranyapin2017evanescent, baaske2014single, young2018quantitative} even in the case that their electric field was uniform. 

Let us take the specific case of Ref.~\onlinecite{mauranyapin2017evanescent}, where the molecule is illuminated by a tightly focused incident field and the scattered light is collected in an optical nanofibre. In such an arrangement the total scattered power is related to the polarizability through Eq.~\ref{scatteredpower}~\cite{alexander2020sounds}. Replacing the term $||\bm{\mu}_{ex}||$ in this equation with its equivalent $||\bm{\alpha}_{ex,k}|| \times ||\bm{E_k}||$ and converting the magnitude of the electric field into scattered optical power gives
\begin{equation}
\langle P_{sc} \rangle = \frac{k^4}{24\pi^2{{\epsilon}_m}^2w^2} \bm{||\alpha_{ex,k}||}^2 \langle P_{in} \rangle,
    \label{scatteredpower_vs_input_power}
\end{equation}
where $w$ and $\langle P_{in} \rangle$ are the width and mean power of the incident field, respectively.

The right axis of Fig.~ \ref{fig5}b shows the scattered power from a  BSA molecule in water undergoing thermal motion at 273~K as a function of time, calculated using Eq.~\ref{scatteredpower_vs_input_power} and normalised to the incident optical power. We assume an optical wavelength of $\lambda = 780$~nm, and that the incident field is focused to a waist with width of $w=1~\mu$m. As can be seen, this is a factor of around $10^{17}$ lower than the input power, meaning that only around one in $10^{17}$ photons will be scattered by the molecule. The scattered power from each conformation of ATPase and proteasome are shown on the right axes in Fig.~\ref{conf_changes}b \textit{(bottom)}. Here, the scattered power is around $10^{16}$ and $10^{15}$ times smaller than the input power, respectively.

Let us consider an experiment that counts the number of scattered photons over some integration period (the temporal resolution of the measurement), and records this photon count as a function of time. To determine whether a conformational change has occurred at a particular time,  one can compare the photon counts  in the integration periods immediately before and after this time. We denote these photon counts $n_a$ and $n_b$, respectively. This approach is illustrated in Fig.~\ref{SNR_conf}. As shown, to resolve a conformational change the mean change in photon count between the two conformation must be larger than the combined statistical fluctuations in the photon counts of the two conformation involved; i.e. the  absolute value of the mean of the measured signal $\langle i \rangle = \langle n_b \rangle - \langle n_a \rangle$ must be larger than its standard deviation $\sqrt{\langle i^2 \rangle - \langle i \rangle^2}$.

A number of noise sources can contribute to the statistical fluctuations, including vibrations, electrical noise, and optical shot noise resulting from the quantisation of light\cite{mauranyapin2017evanescent}. Here we consider only optical shot noise which constitutes a fundamental noise limit in the absence of quantum correlations between photons~\cite{taylor2013biological, casacio2020quantum, taylor2013fundamental}. This limit has been reached in both dark field heterodyne microscopy~\cite{mauranyapin2017evanescent} and interferometric scattering microscopy\cite{Ying2014Shot-noise}. Because uncorrelated photons have Poissonian statistics, the variance of the optical shot noise is equal to the mean number of photons that would be collected in the time interval of the measurement~\cite{taylor2016quantum}. As a consequence the standard deviation is $\sqrt{\langle i^2 \rangle - \langle i \rangle^2} = \sqrt{ \langle n_b \rangle + \langle n_a \rangle}$. The condition to resolve a change in conformation can then be expressed as
\begin{equation}
|\langle n_a \rangle - \langle n_b \rangle| > \sqrt{\langle n_a \rangle + \langle n_b \rangle}.
\label{SNR_conf_changes}
\end{equation}

\begin{figure}[h]
\includegraphics[width=\columnwidth]{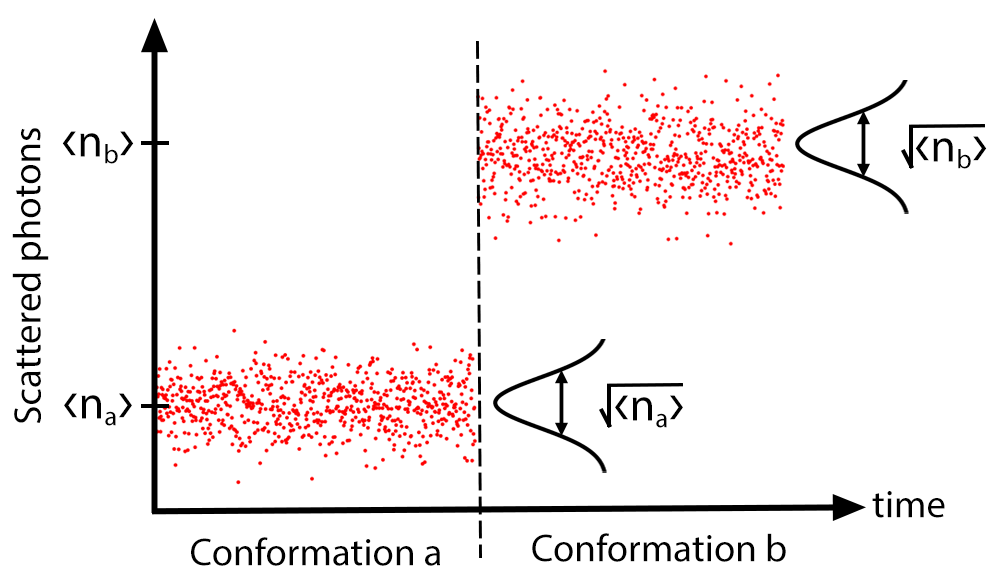}
\caption{\label{SNR_conf} Illustration of the scattered photon flux that might be detected prior to, and after, a conformational change. Here, the random variation in the collected photon number within each period where the conformation is constant can be attributed to optical shot noise, while the discrete jump in the mean scattered photon number can be attributed to the change in polarizability due to the conformational change. Here $\langle n_a \rangle$ and $\langle n_b \rangle$ are the mean scattered photon flux over the measurement time in conformations $a$ and $b$, respectively.  The difference between mean scattered photon number from states $a$ and $b$ must be greater than the noise for the change to be resolvable.}
\end{figure}

The number of photons scattered by a molecule in an integration time $\tau$ is related to the scattered power by
\begin{equation}
\langle n \rangle = \tau \langle P_{sc}\rangle\lambda/hc,
    \label{n_power}
\end{equation}
where $h$ is Planck's constant and $c$ is the speed of light. Using Eqs.~\ref{scatteredpower},~\ref{SNR_conf_changes}~and~\ref{n_power}, and assuming that all scattered photons are collected, the minimum incident optical field intensity needed to resolve a conformational change using a shot noise limited biosensor can be expressed as
\begin{equation}
I_{min} = \frac{48cn_m^2\pi h}{k^4\lambda}\frac{|\bm{\alpha}_{ex,k,a}|^2 \tau_a + ||\bm{\alpha}_{ex,k,b}||^2 \tau_b}{|\hspace{0.5mm}||\bm{\alpha}_{ex,k,a}||^2 \tau_a - ||\bm{\alpha}_{ex,k,b}||^2 \tau_b|^2},
\label{I_min}
\end{equation}
where $n_m$ is the refractive index of the medium, and the subscripts ``a'' and ``b'' are used to label the before and after conformation, respectively. We assume here that the medium is water, with $n_m=1.33$.

Considering the case of ATPase, each conformation exists for a time of roughly 0.7~ms~\cite{nakanishi-matsui2006Stochastic}, so that it is natural to set the integration times to $\tau_a=\tau_b=0.7$~ms. Using, as an example, the polarizability vectors calculated earlier for conformations 1 and 3 when illuminated with an optical field polarized along the \textit{x}-axis (Fig.~\ref{conf_changes}a~{\it bottom}) and using Eq.~\ref{I_min} we find that $I_{min} = 170$~W$/\mu$m$^2$. This intensity is well beyond the range of diffraction-limited single-pass biosensors such as those reported in Ref.~\onlinecite{mauranyapin2017evanescent}, but is comparable with the intensities used in plasmonic and microcavity-based sensors~\cite{mauranyapin2017evanescent, Pang2012Optical, Dantham2013Label-free, Kotnala2014Quantification, knittel2013back}. It is therefore feasible that those sensors, if operating at the shot noise limit, could allow detection of  conformational changes in $\text{F}_1\text{F}_0$ ATPase and related molecules.

For the case of the 26S proteasome, we were unable to source experimental data for the time that the molecule persists in each conformation, though it is known that the molecule reconfigures from conformation s1 through to conformation s3 in around 0.6~seconds\cite{Bard201926S}. Based on this, we assume that each conformation persists for 0.3~seconds, and set an integration time of  $\tau_a=\tau_b=0.3$~s. Comparing the polarization vectors for conformations s3 and s4 when illuminated with a \textit{y}-polarized optical field, we find that these conformations could be distinguished using an optical intensity of $I_{min} = 30$~mW$/\mu$m$^2$. This is within the range of diffraction-limited single-pass biosensors, and is around typical   threshold damage intensities for biological materials\cite{mauranyapin2017evanescent}. We can therefore conclude that proteasome conformational changes could possibly be monitored noninvasively with a shot noise limited optical scattering-based biosensor.

%
%

%


\section{Conclusion}

We have introduced a new efficient method to calculate the dynamic polarizability of macromolecules from their atomic configuration. This method allows the polarizability to be calculated for molecules with size exceeding one hundred thousand atoms. The molecular conformations required as input can be taken from structure repositories such as the Protein Data Bank, or alternatively from molecular dynamics simulations.  By providing a quantitative connection between variations in the polarizability of a protein and its underlying structural dynamics, our method provides a tool that could be used to better interpret single molecule experimental studies of important biological processes such as enzyme catalysis and  protein folding.

\section*{Acknowledgements}
This material is based upon work supported by the Air Force Office of Scientific Research under award number <FA9550-20-1-0391>. It was also supported by the ARC Centre of Excellence for Engineered Quantum Systems (CE110001013). Molecular graphics were produced with UCSF Chimera, developed by the Resource for Biocomputing, Visualization, and Informatics at the University of California, San Francisco, with support from NIH P41-GM103311.

A.M. and S.B. are supported by the ARC discovery program (DP180101421). Computational resources were provided by the Pawsey Centre and the National Computational Infrastructure through the National Computational Merit Allocation Scheme (project m72).

\section*{Author contribution}
L.S.B., W.P.B. and N.P.M. wrote the main manuscript, with input from L.S.M. and E.V.B. L.S.B. and E.V.B. prepared figures. Theory and computational work was Matlab performed by E.V.B. and L.S.B. S.B. and A.M. performed the molecular dynamics simulations for BSA and provided feedback on the manuscript. W.P.B. conceptualized and led the project. All authors approved the final manuscript.

\section*{Competing interests}
The authors declare no competing interests.

\section*{Additional information}
{\bf{Correspondence}} and requests for materials should me addressed to W.P.B.

\section*{References}
\bibliography{ref}

\end{document}